**Critical dynamics mediate stabilization of new distributed memory representations**

Quinton M. Skilling[1], Daniel Maruyama[2], Nicolette Ognjanovski[3], Sara J. Aton[3], Michal Zochowski[1,2]

[1] Biophysics Program, University of Michigan, 930 N University Ave., Ann Arbor, MI 48109

[2] Department of Physics, University of Michigan, 450 Church St, Ann Arbor, MI 48109

[3] Department of Molecular, Cellular, and Developmental Biology, University of Michigan, 830 N University Ave., Ann Arbor, MI 48109

**Corresponding author:**

    Michal Zochowski
    e-mail: michalz@umich.edu

**AUTHOR CONTRIBUTION:** Q.S., D.M., and M.Z. designed and carried out the computational modeling experiments. N.O. and S.J.A. designed and carried out the behavioral and electrophysiological experiments. N.O. and S.J.A. carried out behavioral analyses and spike sorting to discriminate single-unit activity from electrophysiological recordings. D.M. carried out the analysis of network stability on electrophysiological recordings. Q.S. and M.Z. wrote the paper with assistance from N.O. and S.J.A.

**Abstract:**

Critical-state dynamics in the brain have been shown to be an important feature for neural computation. However, links between criticality and network-level mechanisms underlying the formation of new memories are lacking. Here, we record from mice experiencing contextual fear conditioning and probe dynamics for signatures of criticality, finding criticality to be a natural state of the system. We subsequently measure the functional network stability (FuNS) of the recorded neurons and find that tracking changes in the network state before and after fear conditioning accurately predicts fear learning. We turn to modeling to determine the link between critical dynamics and FuNS. We control proximity to a balance between excitation and inhibition, a state we show to be synonymous with criticality, and observe a discrete set of local synaptic changes giving rise to global FuNS only near a critical state. Finally, new information has maximal consolidation potential only near criticality.

**Author Summary**:

Much evidence points to the existence and corresponding implications of self-organized critical states in neuronal systems. Here, we expand on this work and show the importance of criticality on the network sensitivity to input and subsequent consolidation of new memories. Our in vivo studies suggest that critical states provide a necessary substrate for network-wide stabilization of functional interactions between neurons. Through modeling, we provide evidence that this so-called functional network stability is most sensitive only near a critical dynamical state. Further, we show that input has global effects on network dynamics only near criticality and, indeed, that new memories can be formed only at these states. Taken together, these results indicate critical-state dynamics are vital for network sensitivity and consolidation potential.

**INTRODUCTION**

It is widely hypothesized that new information is encoded in brain circuits through activity dependent, long-term synaptic structural changes[1]. These structural changes are a putative mechanism of memory formation[2, 3]. While specific features of memory traces can be localized to specific cell populations (*e.g.*, location information encoded in place cell activity), in general, tracing engrams to specific neural circuits has been an elusive task [2]. Attempts at disrupting well-established memories through brain lesions[4] or, more recently, through optogenetic silencing[5] have shown that they are robust to alterations in communication between individual neurons or brain areas. A parsimonious and longstanding explanation of these phenomena is that a process termed "systems consolidation" leads to diffuse, widespread memory encoding and storage. However, despite more than a century of study, it is not well understood how engrams are initially formed and subsequently stored across vast distances (in terms of numbers of synaptic connections between neurons) in the brain.

A major problem to understanding the mechanisms for systems consolidation is that very little is known about how formation of new memories impacts neural network dynamics. The general, long-accepted assumption is that either strengthening of existing synaptic connections, or the *de novo* creation of additional synapses (*i.e.*, formation of a network-wide discrete structural heterogeneity), leads to the formation of a dynamical attractor[6, 7]. However, the initial number of neurons actively involved in encoding (or recalling) a specific memory trace is thought to constitute a small (on the order of few percent) fraction of the total neuronal population[8]. Moreover, individual synapses in regions like the hippocampus have a surprisingly brief lifetime (~1-2 weeks on average[9]). This raises two questions: 1) how do permanent and widely-distributed neural engrams form from initial, transient changes to a discrete subset of the network's synapses during learning, and 2) what mediates transformation of local representations of disparate features to global memory representation?

Contextual fear conditioning (CFC) is an optimal experimental paradigm to tackle these questions as it allows for rapid formation and consolidation of memory (i.e. after single-trial learning)[10]. Here, we first characterize hippocampal dynamics in mice subjected to CFC and show that: 1) the hippocampus operates in a near critical regime pre- and post- CFC training – a universal dynamical state indicating a dynamic phase transition, and 2) successful consolidation of fear memory leads to stabilization of network-wide functional representations.

While the idea that the brain operates at or near dynamical critically is not new (see for example[11-13] and references therein) the functional benefit of operating in a near-critical regime with respect to the storage of new memories is not clear. We proceed to show computationally that near-critical dynamics in the brain facilitate memory consolidation through increasing functional network stability. Namely, we demonstrate in two classes of reduced models (i.e. mixed excitatory and inhibitory networks near an excitatory/inhibitory (E/I) balance, and in a reduced attractor neuronal network[7]), that **global** changes of representational stability due to **local** structural network modifications, mediated by consolidation of new representations, are universally present near criticality. This is due to the introduction of discrete network heterogeneities (i.e. stronger connectivity within a subgroup of neurons) near critical state leading, universally, to long-range stabilization of temporal network representations.

Finally, we show that near-critical dynamics may be essential for consolidation of new memories in a situation when the sensory input is weak and/or sparse in comparison with signals generated by memories native to the network.

Together, these results indicate that novel learning occurs preferentially near a critical regime and leads to universal widespread stabilization of network activity patterns, which in turn drives the formation of widely-distributed engrams (*i.e.*, systems memory consolidation).

**RESULTS**

**Hippocampal network stabilization *in vivo* predicts effective memory consolidation.**

We hypothesized that synaptic plasticity in hippocampal area CA1 following single-trial contextual fear conditioning (CFC) [14] would be a plausible biological model to investigate how rapid memory formation affects network dynamics. CA1 network activity is necessary for fear memory consolidation in the hours following CFC[15]. For this reason, we recorded the same population of CA1 neurons over a 24h baseline and for 24h following CFC to determine how functional network dynamics were affected by *de novo* memory formation. C57BL/6J mice underwent either CFC (placement into a novel environmental context, followed 2.5 min later by a 0.75 mA foot shock; $n$ = 5 mice), sham conditioning (placement in a novel context without foot shock; Sham; $n$ = 3 mice), or CFC followed by 6h of sleep deprivation (a manipulation known to disrupt fear memory consolidation[16, 17]; SD; $n$ = 5 mice) (**Fig 1A**).

We first characterized hippocampal dynamics recordings over the first 6h of the pre- and post-CFC period (or the first 6h post-SD). Observing the number of active channels during a burst of activity revealed that neuronal dynamics reside near a phase transition (**Fig 1B**). This is exemplified by the fact that the scaling of recorded sizes of bursts of activity (commonly referred to as neuronal avalanches) follow a power-law distribution[11]. We confirmed proximity to critical state dynamics by calculating the goodness of fit, κ, of the observed avalanche distributions to the expected power law distributions near criticality[18, 19] (**Fig 1C**), and 2). Values near unity indicate critical state dynamics, while above or below unity represent supra-critical or subcritical dynamics, respectively. We observe distributions around unity, indicating variability in mouse dynamical state; however, there is no significant difference between mouse groups and all values average out to near one. These findings support the idea that critical-state dynamics are a preferred dynamical state of the brain.

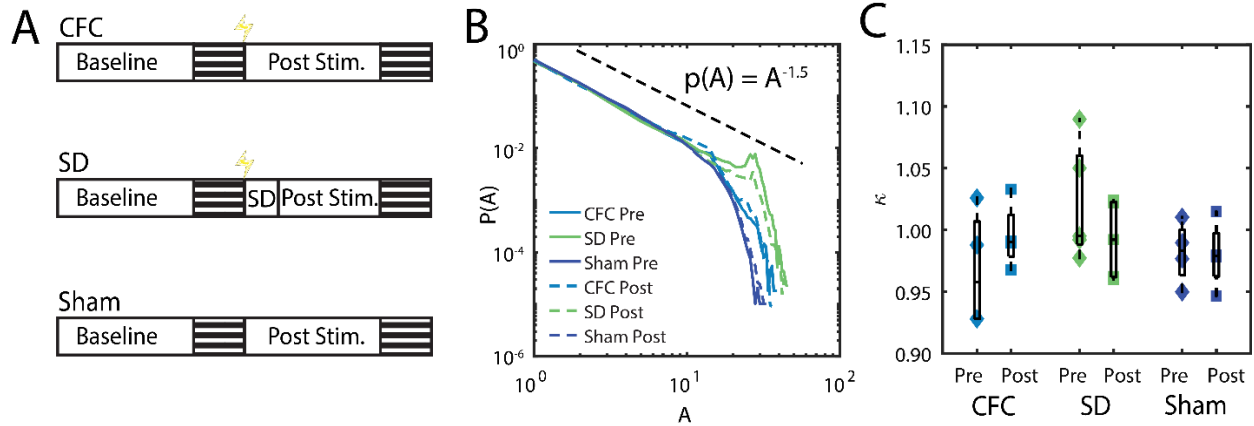

**Fig 1**. Hippocampal LFP recordings indicate near-critical dynamics. **A)** Schematic of experimental procedure. Baseline 24h recordings are taken in a native environment before the mouse is introduced to a novel environment. Mice in the learning groups (CFC and SD) are given a brief foot shock after a brief period of exploration while Sham animals are left to freely explore the novel environment. Follow up recordings are taken when mice are returned to their native environment; SD mice are sleep deprived for 6 hours following foot shock administration whereas CFC mice are allowed *ad lib* sleep over the same period. After subsequent 24h recordings, mice are reintroduced to the novel environment and observed for freezing behavior, an indication of successful consolidation of fear memory. **B)** Spiking events distribute as neuronal avalanches, where the probability of observing a burst of certain size decreases like a power-law. CFC (light blue), SD (green), and Sham (purple) have power-law relationships, both during baseline recordings (solid lines) and after the learning interval (dashed lines). The black dashed line represents a theoretical power-law distribution with exponent -1.5. **C)** Goodness of fit, $\kappa$, calculation for each mouse in respective groups. All $\kappa$ values are reported to be near unity, indicating near-critical dynamics for each group.

We proceeded to investigate why near-critical dynamics maybe a preferred state by characterizing changes in functional network connectivity mediated by rapid formation and consolidation of contextual fear memory (CFM). Here, however rather than reporting pairwise correlations between recorded pairs of cells we set out to measure degree fluctuations or changes in overall functional connectivity in the recorded networks. While the detailed functional connectivity structure provides only local information on interaction between the recorded cells, the degree of overall network changes or lack of thereof (i.e. network stability) provide additional insight about the global network state[20]. To characterize the dynamics of functional connectivity patterns, we implemented a metric we refer to as functional network stability (FuNS; see Methods). Spike trains are divided into temporal bins of the shortest possible length that provides a robust estimate of functional connectivity. Average minimal distance (AMD)[21] is then used to evaluate functional network connectivity between cell pairs within each temporal bin. FuNS is quantified as the average cosine similarity between vectorizations of functional network structures in adjacent temporal bins with values closer to one representing stable (unchanging) functional network structures. The metric is sensitive to rapid reconfiguration of functional connectivity within the network rather than global fluctuations in functional connectivity strength. Therefore, high stability can be detected for weak but unchanging spike correlation patterns – this qualitatively provides information about temporal evolution of functional connectivity as compared to direct assessment of correlation between spike bouts.

We measured change in FuNS in mouse recordings after each manipulation by quantifying FuNS on a minute-by-minute basis over the entire pre- and post-training intervals and calculating their respective difference within each animal. Consistent with previous findings10, we observed an increase in FuNS over the 24 hours following CFC, which was most pronounced when comparing FuNS in slow wave sleep (SWS; **Fig 2A**). In contrast, no change in SWS FuNS was seen in Sham mice or following CFC in SD mice. Group differences in SWS FuNS were reflected in the behavior of the mice 24 hours post-training, when context-specific fear memory was assessed. CFC mice showed an increase in context-specific freezing upon return to the conditioning environment, which was significantly greater than freezing in Sham and SD mice (**Fig 2B**). Consistent with previous results[22], across mice, training-induced changes in SWS-specific FuNS for individual mice were predictive of mouse behavioral performance during memory assessment (**Fig 2C**). Thus, the formation of a behaviorally-accessible memory trace *in vivo* is accompanied by increased FuNS in the CA1 network.

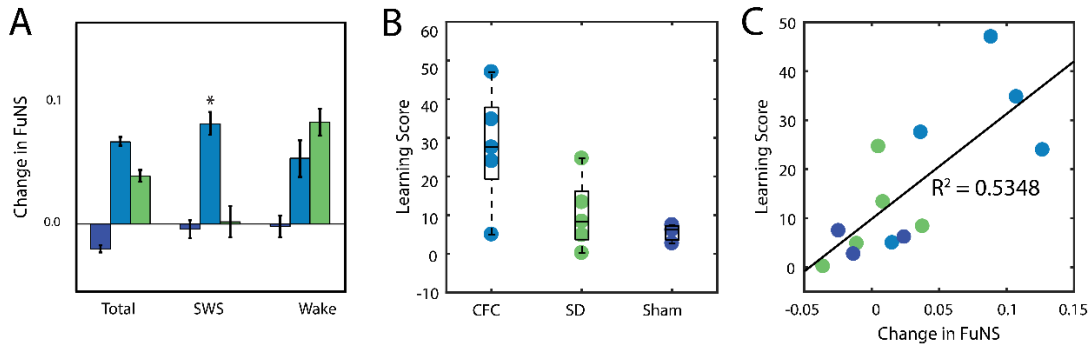

**Fig 2**. Functional Network Stability predicts fear memory consolidation. **A**) Changes in FuNS are tracked across the entire recording (total), during slow-wave sleep only (SWS), and during wakefulness (wake). FuNS in CFC mice show significant separation from the other states and accounts for observed increased stability across the total recording interval. * indicates p < 0.05, Holm-Sidak post hoc test vs. Sham and SD mice. **B**) Distribution of freezing scores for each of CFC, Sleep Dep, and Sham mouse groups. Mice that received a foot shock and were allowed normal sleep patterns afterward exhibited increased freezing behavior over their SD and Sham counterparts. Boxes represent box-and-whisker plots to identify important values of the distribution. **C**) Freezing behavior as a function of FuNS change appears approximately linear. Indeed, fitting a line to the data indicates a goodness of fit of $R^2 \sim 0.53$.

**Local changes in network connectivity universally stabilize network-wide dynamics near criticality.**

Based on our experimental results we hypothesize that during successful CFM formation and consolidation, initial exposure to a stimulus results in rapid, discrete modifications to structural network connectivity, and subsequently affecting network-wide dynamics. We set out to investigate what global impact, and under what conditions, do minor changes in structural connectivity have on stability of global network dynamics. We investigated mixed networks of $N_e$=1000 excitatory and $N_i$=225 inhibitory nodes with sparse connectivity. We changed the relative weights emanating from excitatory and inhibitory nodes and characterized state dynamics as a ratio between excitation and inhibition (E/I) to monitor network response to introduced structural changes. For more information, please refer to the Methods section.

The dynamics around Excitatory/inhibitory (E/I) balance are thought to emerge naturally in neural networks[23]. Neurons operating in networks near E/I balance exhibit faster linear responses to stimulation, and greater dynamic range[23]. Most importantly, it has been shown that that critical regime (phase transition) coincides with E/I balance[18, 19]. Thus, by controlling E/I, we can monitor network dynamics in relation to networks near and far from a critical state.

As above, we characterized network dynamics by evaluating the distribution of avalanche lengths and goodness of fit to an expected power-law distribution (**Fig 3**) to characterize the network dynamics as a function of E/I ratio. We observe near-critical avalanches statistics, κ showing an increase with response to weakening inhibition, crossing from sub- to supra-critical dynamics for high E/I.

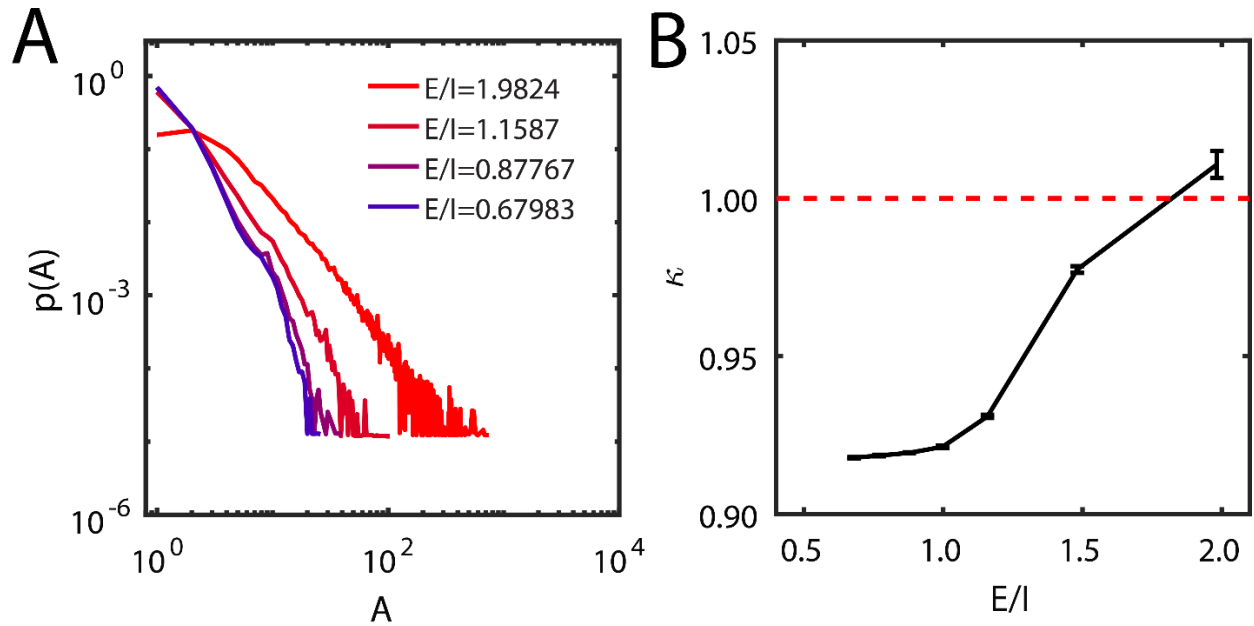

**Fig 3**. Balance between excitation and inhibition controls near-critical dynamics. **A)** Avalanche distributions for different values of E/I ratio in the LIF model. Stronger excitation results in supra-critical dynamics, exemplified by uncharacteristic peaks in observing avalanches of certain size. Stronger inhibition, on the other hand, can be approximated by an exponential decay, indicating sub-critical dynamics. **B)** Goodness of fit, κ, of avalanches in A to a theoretical distribution of exponent -1.5. Avalanche distributions are near criticality, with high excitation resulting in supra-critical dynamics.

We then investigated how addition of a localized network heterogeneity changes the stability of network representations. Here the heterogeneity was defined as strengthened connections between a small (100 neurons) subgroup of adjacently positioned neurons in the network. Functional network stability (FuNS) increased from zero to near unity in a sigmoidal fashion as a function of the E/I ratio (**Fig 4A**). Addition of a synaptic heterogeneity caused a shift in stability dynamics toward lower E/I (red trace in **Fig 4A**), resulting in a maximal change in stability near E/I = 1.2 (**Fig 4B**), precisely the transition site from low to high stability for networks without synaptic heterogeneity (black trace in **Fig 4A**). Most significantly, large deviations of E/I balance in either direction reduce FuNS sensitivity to zero.

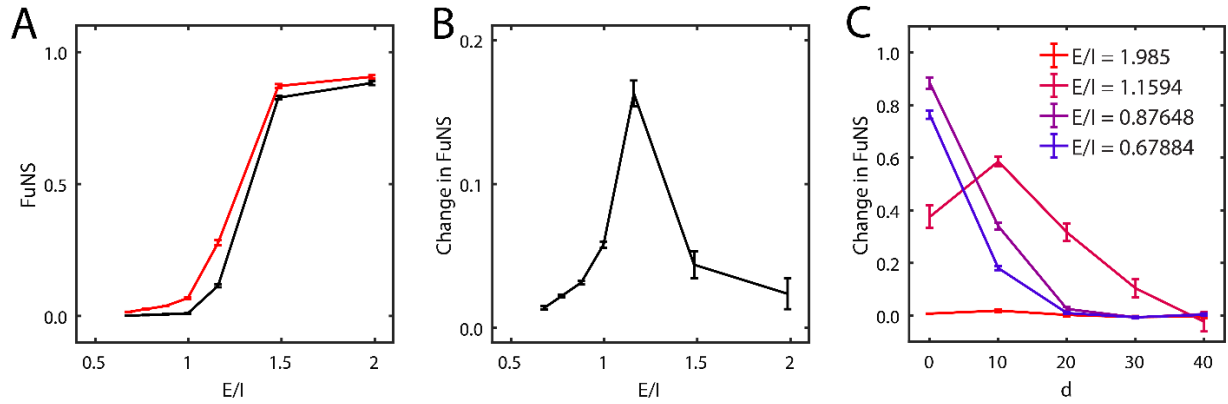

**Fig 4**. Change in FuNS peaks near a balance between excitation and inhibition. **A**) FuNS calculated before (black trace) and after (red trace) introduction of a synaptic heterogeneity, as a function of E/I. **B**) Change in FuNS across the E/I interval. Peak change occurs near a balance between excitation and inhibition, E/I ~ 1.16. **C**) Change in FuNS as a function of mean distance to the synaptic heterogeneity, d. All error bars represent SEM over 10 trials.

To further understand why this is the case we analyzed pairwise network stability changes as a function of network distance between the neurons and the location of heterogeneity. Here we defined the mean network distance as the average number of existing (structural) connections needed for spiking information to traverse from any cell within heterogeneity to the neuron in the question. We observe (**Fig. 4C**) that FuNS changes linearly with distance around criticality, whereas FuNS changes decay exponentially for sub-critical dynamics and remains zero (due to saturation) throughout the network in a supra-critical regime. This result remains consistent with physics of critical points, where correlation length is expected to diverge[24]. Thus, the increase of functional network stability at E/I balance (and thus at critical point) is due to the emergence of global changes in network dynamics after introduction of localized structural heterogeneity. In other words, the network needs to be near criticality for the local heterogeneity to have global effects on dynamics.

**Criticality promotes consolidation of new memories in native networks**

To further understand universal consequences of the above results, we show below using a reduced attractor neuronal network (ANN) [7, 25, 26] that intrinsic properties of the critical state itself may provide distinct dynamical advantages, allowing for storage of new representations from sparse and/or weak input. Namely, we show that introduction of external drive leads to global stabilization of new network representations. Subsequently, when spike-timing dependent plasticity is invoked, this stabilization mediates network-wide storage of the new input representation with the greatest increase in FuNS occurring near a critical transition between order and disorder.

The ANN consists of $N$ = 10000 nodes with binary states $S_i = \pm 1$, arranged in a small-world network (10% chance of rewiring a local connection)[27] with ~2% connectivity. On each integration step, each node changes its state with probability $P(h_i) = \frac{1}{1+\exp(-2\beta|h_i|)}$ to align with its input $h_i = \frac{1}{k}\sum_j^k J_{ij} S_j$, (here, $k$ represents the number of presynaptic nodes and $J_{ij}$ is the weighted connection between the nodes; $\beta = T^{-1}$ is a control parameter which directly controls transition of network dynamics from order to disorder when increased).

We define two respectively uncorrelated fixed states as $\{\xi_i^{e/n} = \pm 1 \mid i \in [1, N]\}$ to represent a natively stored configuration ($\xi^n$) or a new configuration ($\xi^e$) to be stored by the system. We choose a set $M_{inp} = \{1, ..., M\}$ *of M* random nodes from the network to represent persistent external drive of the new configuration by setting $S_{i\,(i \in M_{inp})} = \xi_i^e$ to be constant throughout the simulation (thus impinging its information on the dynamical system). Every connection stemming from $M_{inp}$ nodes has a value $J_{ij} = w^e \xi_i^e \xi_j^e$ ($i \in M_{inp}$ and $j$ is the number of connections) while connections stemming from other nodes have a value $J_{ij} = w^n \xi_i^n \xi_j^n$; $w^{e/n}$ is the weight of a specific configuration. The effect of these strengthened connections is a competition between the local fields emanating from the newly driven state and the native state, stored in the rest of the network.

We measure the network magnetization $m^{n/e}(t) = \left|1/N \sum_{i=1}^{N} S_i(t) \xi_i^{n/e}\right|$ to determine the *overlap* of the network state at time t, with one of the configurations (**Fig 5A**). Below criticality (low T) and for N>>M the steady state of the network achieves a higher overlap with the native state rather than with the new representation. Near criticality (mid T), however, the overlap of the network state with the native state abruptly declines, whereas the network's overlap with the new memory first increases and then attenuates significantly less over this interval, leaving it dominating over the native state. In the supra-critical regime (high T), $m^{n/e} \cong 0$ are the only stable solutions. These results indicate that for system dynamics residing in a subcritical regime, the native representation is a preferential solution for the network state and only near criticality is the fractional overlap with the new representation dominating. This result can also be viewed as diverging magnetic susceptibility at the critical point[24, 25].

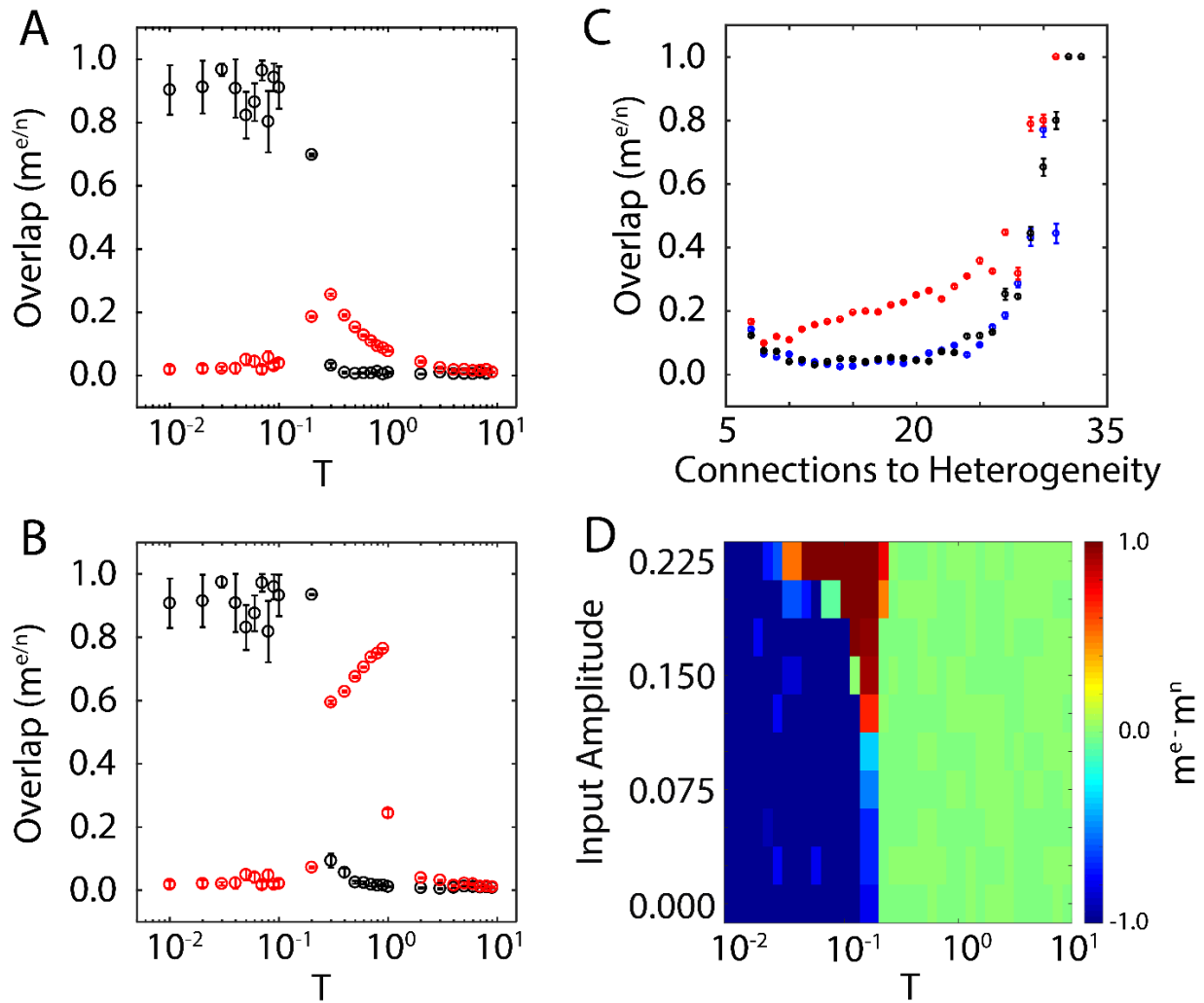

**Fig 5**. New configurations are stored near criticality in an attractor neural network. **A**) Network overlap with a preferred network state (commonly referred to as magnetization by physicists using this model) as a function of T. Native configurations (black points) dominate nodal states for low values of T, with new configurations (red points) dominating only near the phase transition in native configuration overlap. **B**) Overlap as a function of T after synaptic plasticity is introduced in the system. Notably, the new configuration now dominates a majority of nodal states only near the critical region. **C**) Overlap of a subset of the system with a new configuration as a function the number of connections to synaptic heterogeneity experienced by the subset. Black points correspond to a supra-critical regime (T = 5), blue points to a sub-critical regime (T = 0.02), and red points to critical regime (T = 0.3). **D**) Input amplitude as a function of T for constant input field into each node. Color represents difference in overlap between the new and native configurations; negative values (blue colors) indicate dominance of native configurations over new configurations. Similarly, positive values indicate dominance of new over native configurations (red colors). All error bars in panels A-C represent SEM over 10 trials.

We next introduced synaptic plasticity into the system. Connections changed their weights in a state-based manner according to $\Delta J_{ij}(t) = \varepsilon S_i(t) S_j(t)$. Similar to STDP in spiking neurons (Bi and Poo, 2001)[27], this state-based plasticity measure strengthens connections of neurons with similar activity (i.e. state). We found the new representation is stabilized (i.e. $m^e \gg m^n$) only around the critical regime (**Fig 5B**). To further see how the novel network state recruits neural resources near criticality, we calculate the scaling of the overlap $m^e$ as a function of the number of direct connections received from the input neurons for sub-critical, critical, and supra-critical regimes. Here the number of direct connections received from the external input is a proxy for the distance of the node from the external input (see methods). In agreement with **Fig 4C**, the correlation between novel representation and network state decays the slowest around the critical point (**Fig 5C**).

Finally, we investigated stability of novel representations after learning as a function of T and the magnitude of the external field applied during learning. Here, instead of fixing the state of nodes to act as persistent input, each node received a persistent field $h_{\xi^e}$ in addition to the field felt through connections. The color reports the difference between the network overlap with the native and new memory. We observe that for low amplitude, previously applied external fields, the new memory is stable in a narrow window near critical temperature. This region broadens for larger fields (**Fig 5D**).

To further characterize network state dynamics that will allow for direct comparison with that of spiking networks, we measure change of FuNS from before to after learning. We first calculate the functional interaction vector as $\hat{J}(t) = \overrightarrow{S(t)}$, where S(t) is the state spin vector at time t. We then define FuNS as a mean of the dot products of adjacent functional interaction matrices, $FuNS = \frac{1}{n}\sum \frac{<\hat{J}_t|\hat{J}_{t+1}>}{\sqrt{<\hat{J}_t|\hat{J}_t><\hat{J}_{t+1}|\hat{J}_{t+1}>}}$. $FuNS = 1$ when network states remain unchanged and $FuNS \rightarrow 0$ if representations change randomly. The network transitions around the critical point from being inherently stable to highly unstable. During the storage of the novel configuration, due to strengthening of the mean connection strength, the critical point is shifted resulting in significant increase of the network's stability around the critical regime (**Fig 6A-B**). This is not observed away from criticality since in the subcritical region the network is fully stable irrespective of connection strengthening, while in the supra-critical region the network remains unstable.

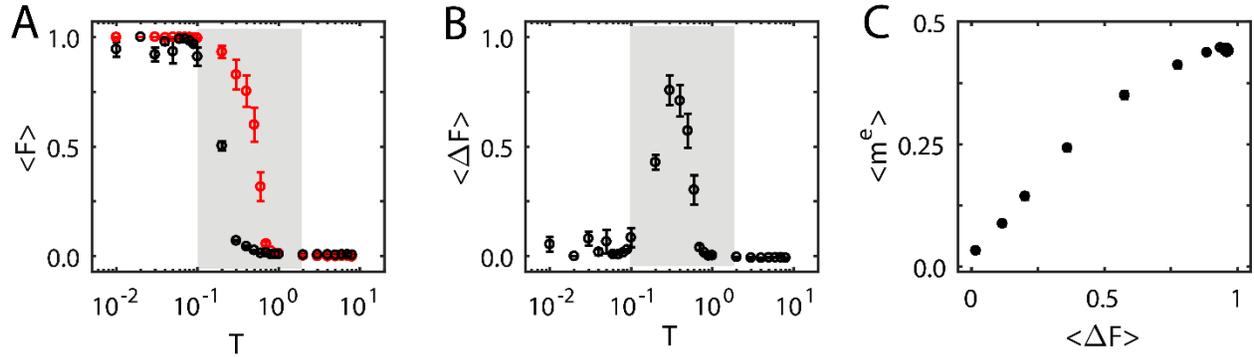

**Fig 6**. Functional network stability scales monotonically with overlap of new configurations. **A**) Functional network stability as a function of T. Black points indicate stability of the native conformation and red points indicate stability of the new conformation when the system is placed in those respective states. **B**) Difference between new and native FuNS curves depicted in (A). Grey panels in A, B represent critical temperature region. **C**) Overlap of the system with the new configuration as a function of stability of the same system over twenty iterations during learning. The relation is approximately linear. All error bars represent SEM over 10 trials.

Finally, we measured the relationship between changes in overlap of a new representation with changes in overall representational stability during learning near criticality, during initial stages of memory consolidation (i.e. the first 20 iterations when synaptic plasticity is activated), (**Fig 6C**). We observe that memory overlap and network stability have a monotonically increasing relationship.

Thus, networks initially near criticality maximally adapt to novel configurations, resulting in increased stability of the system and enhanced learning - this is an important prediction of the model, in line with the results we obtained for experimental recordings (**Fig 2**) and spiking neuron model (**Fig 4**).

**DISCUSSION**

Taken together our results hint at a universal mesoscopic mechanism (functional network stability; FuNS) that underlies what is commonly referred to as "systems consolidation" - *i.e.*, the formation of a widely distributed engram from a transient, discrete and localized group of synaptic changes. We show through *in vivo* experimentation that, while hippocampal dynamics exhibit persistent critical behavior, FuNS is enhanced in CA1 in response to *de novo* memory encoding. Most importantly, the degree of change of FuNS after CFC is predictive of subsequent memory performance. Through computational modeling we show that: 1) FuNS is enhanced throughout a neural network in response to strengthening a discrete subset of network synapses predominantly near critical regime, and 2) critical dynamics may facilitate formation of new memories from weak and/or sparse input.

There is a growing body of work suggesting that the brain self-adjusts its dynamics to operate near criticality[8, 11, 13, 29] – a dynamical state near the point of a second order phase transition, characterized by universal behavior[24]. This state is exemplified by a scaling behavior of activity and the emergence of critical exponents characterizing divergence of systems' observables. While the capacity of the brain to self-organize around this critical point (i.e. the self-organized criticality (SOC) hypothesis) is still being debated[30], work over the past decade has focused on providing experimental support for brain SOC and demonstrating the computational benefits of networks operating near criticality. Regarding the latter, it has been shown that critical dynamics supports optimized information transfer and memory storage capacity (among other network features)[17, 19, 31-35]. More recent studies have shown that neuronal and synaptic responses to external input help neural networks transverse a phase transition toward criticality[36, 37]. Despite this, a unifying mechanistic link between criticality and memory formation is still lacking.

We hypothesize that discrete heterogeneities representing features of a learning experience stabilize functional connectivity within the larger neural network to promote a more distributed memory representation. Widespread network stabilization could promote coordinated consolidation of initially disjointed memory features (*e.g.* location, smell, and color of a rose in a garden), and could later aid in the seemingly instantaneous coordinated recall of these features.

FuNS may mechanistically explain such network phenomena as sequential replay[38] or neuronal reactivation[39] observed experimentally in the context of memory consolidation. For example, highly stable network dynamics would be predicted within neurons forming a strong network heterogeneity, such as might be formed among place cells which are repeatedly activated in sequence during repetitive behavior. The ultimate expression of this stability would take the form of stereotyped sequences of activity, which are highly predictable. Vitally, however, our modeling results predict that neurons several synapses away from the place cell population (possibly in other brain areas) should experience FuNS changes during replay events. Importantly, we present the first data demonstrating that the spatial extent of such network phenomena during post-learning sleep predict successful consolidation.

Our *in vivo* recording data suggest that, contrary to reports of sequential replay or reactivation which typically occur only for ~1h following learning[38, 39], mean change in FuNS in the hippocampus is long-lasting (*i.e.*, across the 24h following training). The spatial extent and duration of FuNS changes are thus highly amenable to promoting systems consolidation of memory. We believe that FuNS could provide a mechanism to drive structural network changes (*e.g.*, through spike timing-dependent plasticity) over widely-distributed networks. If (as recent data suggest) synaptic structures in CA1 are far more transient than synapses in the neocortex[9], the rapid dissemination of memory traces for long-term storage outside the hippocampus is likely required for consolidation. Thus, we would predict that manipulations disrupting FuNS would also disrupt memory consolidation. Indeed, recent data have shown that CA1 network activity alterations that disrupt fear memory consolidation also disrupt post-CFC increases in FuNS[10].

How widespread and universal are FuNS changes after learning? Our computational data suggest that the level of FuNS increase at any given distance from a network heterogeneity is proportional to both the degree of synaptic strengthening and the number of synapses strengthened. Thus, the area over which the network is stabilized would be proportional to the strength of encoding. However, it is highly likely that new memory traces initially affecting hippocampal circuits would alter FuNS among neocortical neurons just a few synapses away. Previous data have indicated that novel experience in wakefulness leads to subsequent "reverberation" of ensemble activity in subsequent sleep; this reverberation persists over hours to days and is present across numerous distant brain areas[40]. FuNS would provide an explanatory mechanism for this phenomenon, which is based on the persistence of spike timing relationships established during experience. Our in vivo and in silico models predict that evolutionarily-conserved features of brain networks promote FuNS increases in response to learning, indicating that FuNS may prove to be a universal feature underlying systems memory consolidation on the mesoscopic scale, (i.e. when measured relationships between networks of individual neurons). Here, our findings are complementary to earlier results that showed that on the macroscopic scale (i.e fMRI measurement of brain regions) there is an increase of flexibility of functional connectivity on short timescales that decreases during late phases of learning[41, 42]. Thus, future experiments will be needed to determine whether FuNS changes are also present following forms of learning that are dependent on synaptic depression rather than potentiation, and that are independent of the hippocampus.

**METHODS**

## 1. Computer simulations

### 1.1. Leaky Integrate and Fire Model

Networks were composed of leaky integrate and fire (LIF) neurons with individual voltage dynamics governed by $C\frac{dV_i}{dt} = -\alpha V_i + I_{syn} + I_{const} + I_{noise}$. Here, each neuron is noise driven, with leak $\alpha = 0.2$, constant current $I_{const} = 0.15$, and noise $I_{noise} = 10$ with a probability of $p = 10^{-4}$ and $I_{noise} = 0$ otherwise. The synaptic current is given as the sum of all synaptic inputs into the cell, $I_{syn} = \sum_j \omega_{ij} \left( e^{-(t-t_j^{spk})/3} - e^{-(t-t_j^{spk})/0.3} \right)$, where $t_j^{spk}$ is the time of the last spike in neuron *j* and the double exponential represents the pulse shape of an action potential. The synaptic weight $\omega_{ij}$ is determined by the cell type of the output neuron *j*; $\omega_{ij} = 0.07$ if the output neuron is excitatory and $\omega_{ij} = -0.07\beta$ if the output neuron is inhibitory. The constant $\beta$ is a control parameter used to vary the ratio between excitation and inhibition in the system, E/I.

LIF networks were composed of a total of 1,225 neurons arranged in a ring network, with 225 inhibitory neurons equally spaced throughout the network. Each neuron was connected locally to 2% of the network, resulting in ~24 connections per neuron. A continuous region of synaptic heterogeneity was introduced comprising 100 neurons. Synaptic connections between neurons within the heterogeneity were doubled without regard to excitatory or inhibitory classification. All simulations were conducted over 10 trials.

### 1.2. Dynamics and memory storage in an auto-associative memory model

A two-dimensional square network comprising $N + M$ = 10,000 nodes is simulated. Each node is binary with state given by $S_i = \pm 1$. A subset $M_{inp} = \{1, ..., M\}$ of these nodes are chosen to represent an external input, which is fully represented by the state $\{\xi_1^e, \xi_2^e, ..., \xi_{N+M}^e\}$, separate from the native state $\{\xi_1^n, \xi_2^n, ..., \xi_{N+M}^n\}$, and are not allowed to change throughout the simulation; the remaining N neurons change normally based on a field-based probability as detailed below. Each node is connected to ~2% of the network, where connections are predominantly local with a probability of $p = 0.1$ being rewired to a random target. The strength of each connection is defined by $J_{ij} = w^{e/n}\xi_i^{e/n}\xi_1^{e/n}$ with the superscripts given by e (external input) only if $j \in M_{inp}$ and n (native) otherwise; here $w$ is a weight value representing the relative strength of the external input to the native state. The total input to each node is then given by $h_i = \frac{1}{j}\sum_j J_{ij} S_j$, where the summation is over all j connections. Finally, during each step of the simulation, a node aligns to the sign of its input with probability $P(h) = \frac{1}{1+\exp(-2\beta|h|)}$ where $\beta = T^{-1}$ controls the dynamic regime (subcritical, critical, and supra-critical for low, medium, and high T, respectively). All simulations are initialized to random conditions. Unless otherwise stated, all figures are results for an external input of M = 700 neurons and a weight ratio given by $\frac{w^e}{w^n} \sim 3.5$.

A learning rule was implemented to facilitate the consolidation of a new representation to memory. The form of plasticity is generic and state dependent, resembling spike timing dependent plasticity (STDP) in spiking systems[28]. During each time step in the learning phase of the simulation, the synaptic connectivity between all nodes changes by some small amount following $J_{ij} = J_{ij} + \varepsilon S_i S_j$ where ε= 0.1.

All simulations were conducted over 10 trials.

### 2. Statistical Methods

#### 2.1. Avalanche Detection

Neuronal avalanches are defined as unbroken temporal clusters of neural activity bounded by periods of quiescence. Avalanche statistics can be elucidated from either discrete or continuous signals, e.g. from spiking data in computational simulations or from recordings of the oscillatory local field potential.

From hippocampal mouse LFP recordings, we calculate avalanches following previous work by others[11]. In short, we integrate the area under the curve of negative deflections in LFP and identify enough events, through thresholding, to satisfy an expected frequency of activity over the recording interval, thereby creating a binary vector of event times for each recording site[11, 43]. Similarly, discrete timing of events (i.e. spikes produced through modeling) directly populate a binary activity vector for each cell spanning the length of the simulation.

Neuronal avalanche statistics are then calculated by counting the total number of active channels during a burst of consistent temporal activity; the size of the avalanche is the number of channels participating in the activity[44]. Due to the high sampling frequency of recorded LFP data, bin sizes equal to the mean inter-event interval were used during calculation of avalanche statistics; no binning was necessary for discrete-timed Integrate-and-Fire data.

## 2.2. κ index to determine proximity to criticality

Avalanche distributions are analyzed for proximity to criticality using the κ Index[18, 19]. The κ index is determined by comparing the cumulative distribution function of calculated neuronal avalanches to the cumulative distribution function of a power-law distribution with an exponent of -1.5 at equidistance avalanche sizes. Specifically, the κ Index is given as one minus the average difference between the calculated CDF and the expected CDF: $\kappa = 1 - \left(\frac{1}{n}\sum_i^n CDF_i^{theory} - CDF_i^{calculated}\right)$ where $n$ is the number of equidistant points considered. System dynamics as reported by κ can either be sub-critical (κ < 1), critical (κ = 1), or supra-critical (κ > 1).

### 2.3. Calculation of pairwise synaptic distance within the networks

The synaptic distance *d* between two distinct nodes in a network indicates the path length separating nodes along their connections. The most direct way to determine $d_{ij}$ for each node in the network is to calculate $A^n$ (the matrix product of the adjacency matrix with itself n times) for sufficient *n,* until all pairwise connections are determined[45]. The distance $d_{ij}$ assumes the first value of *n* which renders a switch from zero to a non-zero value for the pairwise interaction. The diagonal of subsequent matrix products should be set to zero to ensure no self-connections as these values yield spurious results.

Small-world networks, by nature, result in a reduced path length between any two nodes in the system[27]. In this case, the average synaptic distance from heterogeneity to a given site is measured as the average number of connections from every neuron constituting heterogeneity to the site in question.

### 2.4. Average minimal distance and Functional Network Stability

Average Minimal Distance (*AMD*)[21] was applied to network spiking data to determine functional connectivity. *AMD* calculates the mean value of the smallest temporal difference between all spikes in one neuron and all spikes in another neuron. Analytical calculations of the expected mean and standard deviation of minimal distance is then used to rapidly determine the significance of pairwise minimal distance. Specifically, the first and second raw moments of minimal distance for each node are calculated: $\mu_1^L = \frac{L}{4}$ and $\mu_2^L = \frac{L^2}{12}$, where L is the temporal length of the interspike interval and we have assumed that (looking both forward and backward in time) the maximum temporal distance between spikes is $\frac{L}{2}$. Over the entire recording interval *T*, the probability of observing an interspike interval of length L is simply $p(L) = \frac{L}{T}$. Then, the first and second moments of minimal distance considering the full recording interval are given as $\mu_1 = $ $\frac{1}{4T}\sum_L L^2$ and $\mu_2 = \frac{1}{12T}\sum_L L^3$. Finally, the calculated statistical moments give rise to the expected mean and standard deviation, $\mu = \mu_1$ and $\sigma = \sqrt{\mu_2 - \mu_1^2}$, which are used to determine the Z-score significance of pairwise connectivity: $Z = \sqrt{n}\frac{AMD_{ij} - \mu_i}{\sigma_i}$. Values of $Z_{ij} \geq 2$ represent significant functional connections between node pairs.

Functional Network Stability (FuNS) tracks global changes in network functional connectivity by quantifying similarities in *AMD* matrices over a recording interval. The procedure is as follows: first, a recording interval is split into *n* partitions of equal temporal length. Each partition is subjected to *AMD* functional connectivity analysis, resulting in *n* functional connectivity matrices *Z*. Similarity between time-adjacent functional networks is determined using the normalized dot product after matrix vectorization. FuNS is then determined by taking the mean of these cosine similarities over the recording interval: $\text{FuNS} = \frac{1}{n-1}\sum_{t=1}^{n-1}\frac{<Z_t|Z_{t+1}>}{||Z_t||\,||Z_{t+1}||}$. Thus, not only does FuNS yield insight into how functional connectivity changes over time, it can also shed light on how behavior affects the underlying functional network, e.g. by quantifying the difference between FuNS calculated before and after a learning task.

### 3. *Chronic in vivo recording and contextual fear conditioning*

Male C57BL6/J mice (Jackson, aged 2-5 months) were implanted with driveable headstages containing two bundles of 7 stereotrodes each (spaced 1 mm apart) for single-unit and local field potential (LFP), and silver-plated wires for nuchal electromyographic (EMG) recording. LFP and EMG signals were used to assign behavioral states (wake, SWS, REM sleep) in 5-s epochs throughout the recording period. Mice were individually housed (in standard caging with beneficial environmental enrichment including nesting material, manipulanda, and treats) during post-operative recovery and subsequent behavioral experiments. Lights were maintained on a 12hr:12hr light:dark cycle, and food and water were available *ad lib*, throughout all procedures. All housing and experimental procedures were approved by the University Committee on Use and Care of Animals at the University of Michigan.

Following a 1-week recovery period, mice were habituated to daily handling (5-10 min/day) for 3 days. During this habituation period, stereotrodes were gradually lowered into CA1 until stable neuronal recordings (with characteristic spike waveforms continuously present on individual recording channels for more than 24 hr) were obtained. After this, no changes to electrode position were made throughout subsequent experimental procedures. All mice underwent a 24-hr baseline recording starting at lights on. At lights on the following day, mice underwent single-trial contextual fear conditioning (CFC) or sham conditioning (Sham, $n$ = 4)[10]. Mice were placed into a standard conditioning chamber (Med Associates) with patterned Plexiglass walls and a metal grid floor. All mice were allowed to freely explore the novel chamber over the 3-min training session; CFC mice (but not sham mice) received a 2-s footshock (0.75 mA) after the first 2.5 min. At the end of 3 min in the conditioning chamber, mice were returned to their home cage for a 24 hr post-conditioning recording period. CFC mice were subdivided into two groups - one which was allowed *ad lib* sleep (CFC, $n$ = 5), and a second which was sleep deprived by gentle handling for the first six hours following training (a manipulation which is sufficient to disrupt contextual fear memory consolidation[15-17]; SD, $n$ = 5). 24 hour following training, at lights on, mice were returned to the conditioning chamber for a 5-min assessment of contextual fear memory. This was calculated as the change in context-specific freezing between testing and training trials (i.e., % time spent freezing at test — % time spent freezing at baseline [pre-shock]).

Electrophysiological signals were digitized and differentially filtered as spike and LFP data as described previously[10] using Omniplex hardware and software; single-unit spike data was discriminated using Offline Sorter software (Plexon). The firing of individual neurons was tracked throughout each experiment on the basis of spike waveform, relative spike amplitude on the two stereotrode recording channels, positioning of spike wave-form clusters in three-dimensional principal component space, and neuronal subclass (e.g., FS interneurons vs. principal neurons). Only those neurons that were reliably discriminated and continuously recorded across 24-hr baseline and 24-hr post-conditioning recording periods were included in analyses of network stability.